# On the possible common nature of the ground state in Cu- and Fe-based HTSCs

*K.V. Mitsen[1] and O.M. Ivanenko*

*Lebedev Physical Institute RAS, 119991 Moscow, Russia*

**A qualitative model describing the ground state and the mechanism of superconducting pairing in Cu- and Fe-based high-temperature superconductors (HTSCs) is suggested. In this model, doping by localized charges (as well as physical or chemical pressure) is supposed to be responsible for transition of Cu- and Fe-based HTSC to new ground state common for both HTSC classes where specific mechanism of superconductivity takes place. The resulting HTSC ground state is strongly correlated insulator with not fully filled exciton-electronic band, where the incoherent electron transport is impossible but coherent superconducting transport is possible because the band is not fully occupied. It is shown also that such electronic system is inherently predisposed to superconductive pairing because each pair of nearest cations acts as a two-atom negative-U center. The nature of Fermi arcs and mechanism of pseudogap are considered. It is shown that both of these features result from *d*-wave pairing and therefore have to be observed only in cuprates. We believe that the considered ground state is common for various families of HTSCs including cuprates, pnictides, selenides, bismutates and probably some other.**

---

[1] E-mail: mitsen@sci.lebedev.ru

**Introduction**

The nature of the normal state and the mechanism of superconductivity in Cu- and Fe-based HTSCs remain a matter of intense discussions. It is commonly agreed that, in Fe-based HTSCs, only Fe states appear at the Fermi level and it is these states that are responsible for transport, galvanomagnetic, and superconducting properties of these compounds. This means that strong electron correlations at the cation, which determine the band structure of cuprates, are absent in Fe-based HTSCs.

However, in spite of these differences, it looks like in both cases there exists some general and fairly "coarse" mechanism independent of fine details of the band structure and responsible for superconductive pairing in these materials.

Guided by this idea, we will concentrate on the distinctive features common to cuprates and pnictides. Such features do exist and are listed below.

- Low concentration of the charge carriers. Even at the optimum doping, the carrier density is lower than $10^{22}$ cm$^{-3}$; thus, the average distance between the carriers $r_s >$ 3Å, which exceeds the anion−cation spacing. This means that the interaction within a unit cell is essentially unscreened, which makes possible the existence of well-defined electron−hole excitations (charge-transfer excitons) [1].

- The presence of quasi-two-dimensional layers of 3d metal cations ($Cu^{2+}$, $Fe^{2+}$) and ligand anions ($O^{2-}$, $As^{3-}$). This structure stipulates a large contribution of the bulk Madelung energy $E_M$ to the electronic structure of anion−cation planes. It is possible to control the local value of $E_M$ by local electron- or hole doping.

- The existence of an interband gap $\Delta_{ib} \sim$ 2 eV between the occupied anionic band and unoccupied states of the cation band in the electronic spectra of undoped materials.

With both families being high-temperature superconductors, it seems reasonable to relate this fact to features they have in common. In this we will propose a qualitative model for which these features play the key role. In this model, doping by localized charges (as well as physical or chemical pressure) is responsible for the suppression of the gap between the occupied anionic band and unoccupied states of the cation band and the formation of an electron-excitonic band of unusual nature. It is shown that the resulting electronic structure is favourable for the realization of a peculiar mechanism of the electron–electron interaction. It is shown that the proposed model enables explanation of a number of features in both Cu- and Fe-based HTSCs.

## 1. Model of the HTSC electronic structure

As was mentioned above, in undoped cuprates as well as in undoped pnictides the transfer of an electron from an anion to a cation requires an energy $\Delta_{ib}$ (Figs. 1a, 1d). In the case of cuprates (Fig. 1a), it is thought that $\Delta_{ib}$ is related to the Coulomb correlation of electrons at Cu ions; in the case of Fe-based HTSCs (Fig. 1d), $\Delta_{ib}$ is the band gap. However, in both cases it is possible to control the value of $\Delta_{ib}$ by physical (or chemical) pressure, or by doping (locally decreasing the Madelung energy)[2].

Suppose that $\Delta_{ib}$ is decreased for one of the reasons cited above. If this decrease is sufficiently deep so that $\Delta_{ib}$ vanishes altogether, new bands, formed by hybridized states of d and L bands, appear at the Fermi level (Figs. 1b, 1e).

The resulting new conduction band can be either half-occupied (as in cuprates (Fig. 1b), with odd number of electrons per unit cell) or partially occupied (as in pnictides (Fig. 1e), with even number of electrons per unit cell under the condition of band overlap). If the reducing of $\Delta_{ib}$ was achieved by doping of holes (electrons) in cation-anion plane, this material will exhibit electron (hole) conduction in the first case (Fig. 1c) and hole (electron) conduction in the second case (Fig. 1f). This is in full agreement with the types of conduction observed in overdoped phases of Cu- and Fe-based HTSCs upon hole and electron doping.

In both cases, an excitation with the energy $\Delta_{ib}$ corresponds to the transfer of an electron from an anion to a cation band with the formation of a band hole. However, it is another, exciton-like excitation $3d^{n+1}(L^-)$, which has the lowest energy $\Delta_{ct} < \Delta_{ib}$. This excitation corresponds to the transfer of an electron from an anion (O, As) to the nearest 3d cation (Cu, Fe) with the formation of an $L^-$ hole localized at the neighboring anions (Figs. 1g, 1i). Thus, as $\Delta_{ib}$ are gradually reduced, one expects that a state with $\Delta_{ct} = 0$ be attained first.

---
[2] An explanation of doping mechanism will be proposed below.

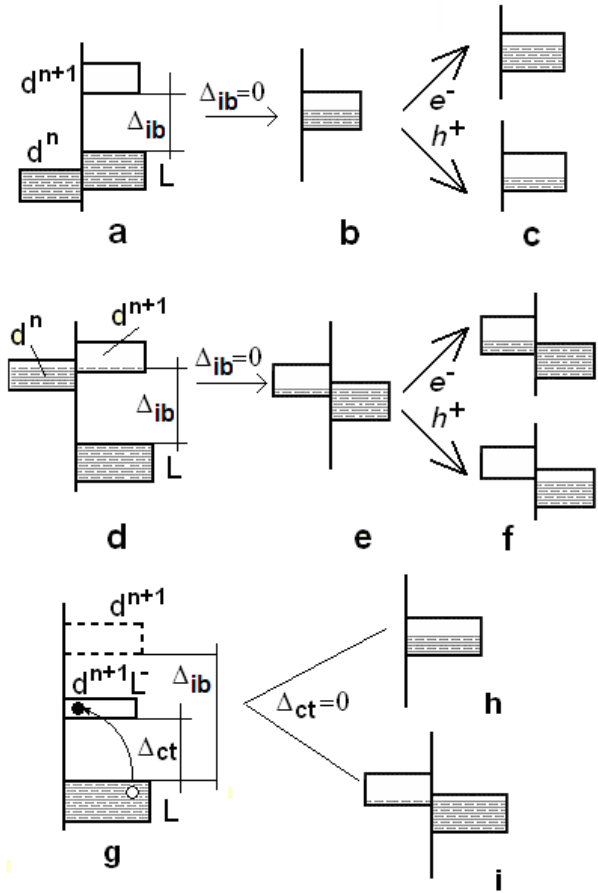

**Figure 1**. Modification of the electron structure of Cu- and Fe-based HTSCs upon a decrease in the interband gap $\Delta_{ib}$. a, d – the electronic structures of undoped cuprates and pnictides, respectively; b, e – modifications of the electron spectra "a" and "d" with $\Delta_{ib}$ somehow reduced to zero; c, f – the shape assumed by the electron spectra "a" and "d" as $\Delta_{ib}$ vanishes owing to the electron or hole doping; g – the minimal energy for the interband excitation in cuprates and pnictides is the energy of the exciton-like excitation $\Delta_{ct}$, which corresponds to the transfer of an electron from an anion to a neighboring cation with the formation of a localized hole (for pnictides, the dashed rectangle denotes unoccupied states in the Fe3d band); h,i – modifications of the electron spectra "a" and "d" for vanishing $\Delta_{ct}$.

Suppose that a state with $\Delta_{ct} = 0$ for the entire anion−cation plane is attained with no extra carriers added to the plane. The condition $\Delta_{ct} = 0$ implies that $3d^{n+1}(L^-)$ and $(3d^n)L$ states are on the chemical potential level. In this case the state of the system can be considered as a state with one half-occupied band for cuprates (Fig. 1h) and two overlapping bands for pnictides (Fig. 1i).

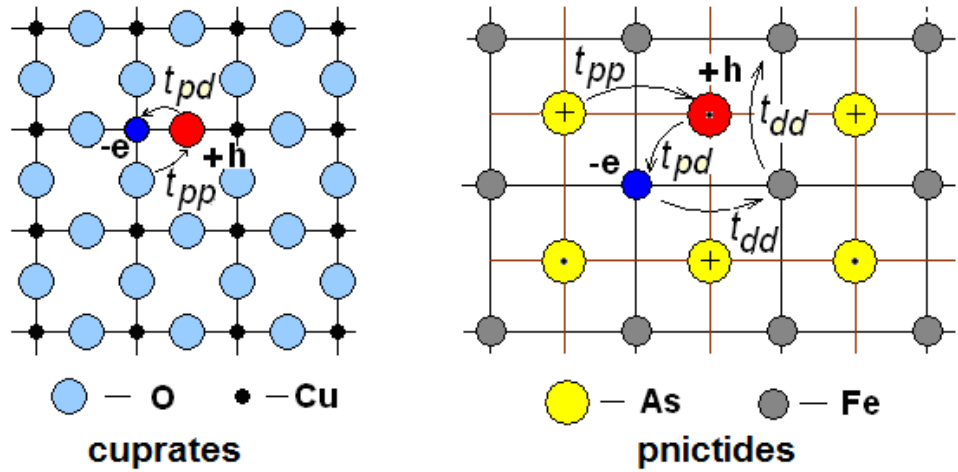

**Figure 2**. Possible transitions of electrons in the anion-cation plane for $\Delta_{ct}= 0$: (a) cuprates, (b) Fe-based HTSCs.

At the same time, the transition of an electron to a cation is possible only in the presence of a hole on one of the nearest anions (Fig. 2a,b). Therefore, an electron and hole can only move around each other or together (as exciton). This means that incoherent electronic transport cannot occur in this system.

However, despite the fact that incoherent electronic transport in such a system cannot take place, the existence of a FS leads to the possibility of coherent transport, when the entire electron system moves as a whole (a condensate). Besides, incoherent hole transport is possible in this system if there were a mechanism of free-hole generation. Below we discuss both of these possibilities.

## 2. The superconducting pairing and normal transport

Let us now examine a possible mechanism whereby a coherent superconducting state may be established. We will show that the systems under consideration are inherently predisposed to superconductive pairing because each pair of nearest cations acts as a two-atom negative-U center (NUC) [2].

As follows from the previous section the transition of an electron to a cation at $\Delta_{ct} = 0$ is possible only in the presence of a hole on one of the nearest anions . At the same time the localization of one extra electron on $d$-cation reduces the charge of the cation by 1. If two electrons are sited on neighboring $d$-cations the Coulomb repulsion between these cations decreases on the value of ~ 1.8 eV (for cuprates without considering of screening). This means that each pair of neighboring cations in such a system can be regarded as NUC on which the pair of electrons has a negative correlation energy and can form a bound state. It is this interaction we believe to be responsible for the superconducting pairing mechanism in these materials. Electron pairing in *k*-space can be considered as a result of the exchange by local excitons [1]. Meanwhile electron transfer in this system can take place at $T<T_c$ owing to the transition to coherent superconducting state at some $T = T_c$, which is governed by $|\Delta_{sc}(\boldsymbol{k})|$.

As noted above, at $T>T_c$ electron transport is impossible, however hole transport could be realized provided that there is some mechanism for the generation of free holes. We show that the mechanism of hole generation could be realized in these materials.

It is obvious that electron energy lowering under electron transition to NUC is enough to close the interband gap $\Delta_{ib}$ and the transition of two electrons to NUC at $T>T_c$ must be accompanied by the appearance of two free holes. Just these holes (not doped carriers, which are localized!) as we believe provide incoherent transport in normal state.

In ARPES the system under consideration will show Fermi surface, whose parameters are determined by the hopping integrals $t_{pp}$, $t_{pd}$ and $t_{dd}$ (Fig. 2). For Cu-HTSC it will be single Fermi surface (FS), and for Fe-HTSC there will be two Fermi surfaces (electron- and hole-like) with equal concentrations (without doped charges consideration). Note that the processes of hole transport (at $T > T_c$) and coherent electron transport (at $T < T_c$) will be characterized by opposite signs of the carriers.

## 3. Symmetry of the order parameter

Because the electrons attract each other being on the neighboring cations, the *k* dependence of the order parameter (or pairing potential) is determined by the *k* dependence of the rate of electron transitions to the cation sites.

Since there is no overlap of Cu orbitals in cuprates, electrons moving in the diagonal direction (along the O−O bonds) cannot appear at the Cu sites. At the same time, the highest transition rate is expected for directions along the Cu−O bonds. For any direction of the wave vector, the rate of electron transitions to the Cu sites is proportional to the length of the OA segment (Fig. 3). It is easy to see from Fig. 3 that AB=cos($k_x a$)-cos($k_y a$) for the unit circle. Thus, the order parameter in cuprates has d-wave symmetry.

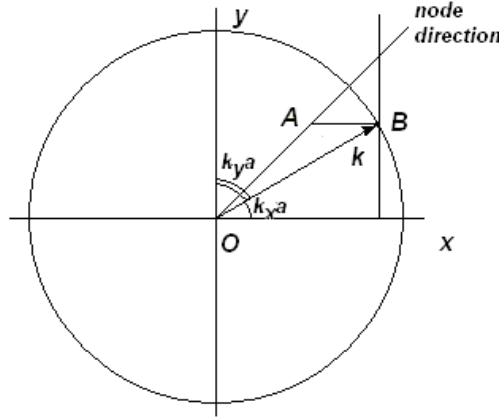

**Figure 3**. Evaluation of the order parameter in cuprates. The rate of electron transitions to the Cu sites is proportional to the length of the segment AB=cos($k_x a$)-cos($k_y a$).

In Fe-based HTSCs because of overlapping of Fe orbitals the order parameter will have no nodes and will have s-symmetry.

### 4. Two mechanisms of doping of parent HTSC phases

Now we consider mechanisms responsible for the modification of the electronic structure of HTSCs by heterovalency doping. Within the simplest ionic model, the gap value $\Delta$ for the transfer of an electron from an anion to the nearest cation is given by the following formula:

$$\Delta \sim \Delta E_M + A_p - I_d$$

Here $I_d$ is the ionization potential of cation, $A_p$ is the electronegativity of anion, and $\Delta E_M$ is the difference in the electrostatic Madelung energies between two configurations, in which the charges of neighboring cation and anion are changed on ±1. It is possible to control the local value of $\Delta$ by local electron or hole doping changing locally the Madelung energy. What is important is that adding electrons (to cation orbitals) or holes (to anion orbitals) leads to the same result: a decrease in $\Delta E_M$ and, hence, in $\Delta$. We will suppose that charges (both holes and electrons) introduced by doping into HTSC parent phases are localized in the nearest vicinity of the dopant ions[3]. The interaction between cation and anion within a unit cell is supposed to be unscreened.

Taking into account the localization of doped carriers we will divide all Cu- and Fe-based HTSCs into two classes: "noble" HTSCs, where doped charges are localized outside the anion−cation planes, and "ignoble" HTSCs, where doped charges are localized in the anion−cation planes.

The first class includes YBCO, BSCCO, and some others. The role of doped charges in these HTSCs consists in the closing of $\Delta_{ct}$ gap in their nearest vicinity. As an example, Fig. 4 shows the mechanism of doping in YBCO. A doped hole (+$e$) from the excess oxygen ion in the chain is distributed over 4 oxygen ions in the vertical copper-oxygen plaquette. The presence of a positive charge (~+$e$/4) on the apical oxygen nearest to a Cu ion in the $CuO_2$ plane is enough to close the gap $\Delta_{ct}$ for electron transitions to this Cu ion from neighboring oxygens. However, the charge of +$e$/4 is insufficient to close $\Delta_{ib}$.

The second class (Fig. 5) includes all HTSCs where charges (+$e$ or −$e$) are introduced directly into anion−cation planes. In particular, this class includes LSCO, NCCO, and some other Cu- , as well as Fe-based HTSCs.

For hole doping, an extra doped hole is distributed by symmetry over 4 anions around a central cation. The total value of such charge (+$e$) is enough for closing of the $\Delta_{ib}$ gap for electron transitions to the central cation from neighboring anions. This area inside this anion square represents a metallic nano-islet (or overdoped islet). At the same time the value of charges of bordering anions (+$e$/4) is enough to close the gap $\Delta_{ct}$ for electron transitions to nearest cations from neighboring anions. As a result, a one-cell layer with $\Delta_{ct}=0$ is formed around such an overdoped islet.

For electron doping, the extra doped electrons (-$e$, or -2$e$ for Co-doped pnictides) are distributed by symmetry over several cations (5 or more) around a central cation, i.e. occupy cation orbitals. The border of the electron localization area is defined by the condition that the charge of boundary ions is sufficient for closing $\Delta_{ct}$ in one-cell layer formed around the electron localization area. The total amount of extra charge inside the localization area is enough for closing the $\Delta_{ib}$ gap for transitions of electrons to cations from anions. This area represents a metallic nano-islet (or overdoped islet) with $\Delta_{ib}=0$ just as in hole-doped HTSCs.

---

[3] We speculate that localization is the result of self-trapping of doped charges caused by a local modification of the electronic structure of anion-cation plane, which consists in the formation of a metal nano-islet whose size is limited by an arising barrier with $\Delta_{ct}=0$.

Therefore, around metallic nano-islets with $\Delta_{ib}=0$ in the anion-cation plane, there forms a layer with $\Delta_{ct}=0$. As the doping level increases (near the threshold of percolation through metallic nano-islets), extended clusters with $\Delta_{ct}=0$ are formed (Fig. 5).

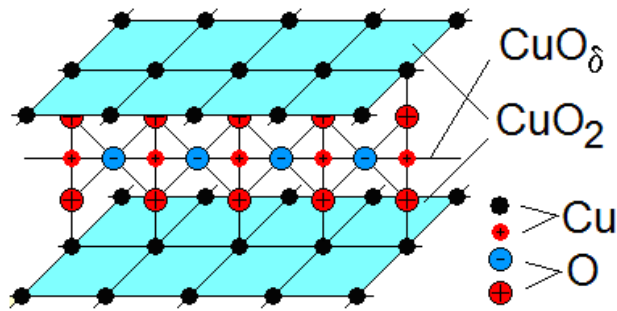

**Figure 4**. The mechanism of doping in YBCO. Oxygen doping results in appearance of localized positive and negative charges in $CuO_\delta$ chain. These charges close the $\Delta_{ct}$ for the electron transition between oxygen anions and Cu cations in two in the two $CuO_2$ planes. Oxygen ions in $CuO_2$ planes are not shown.

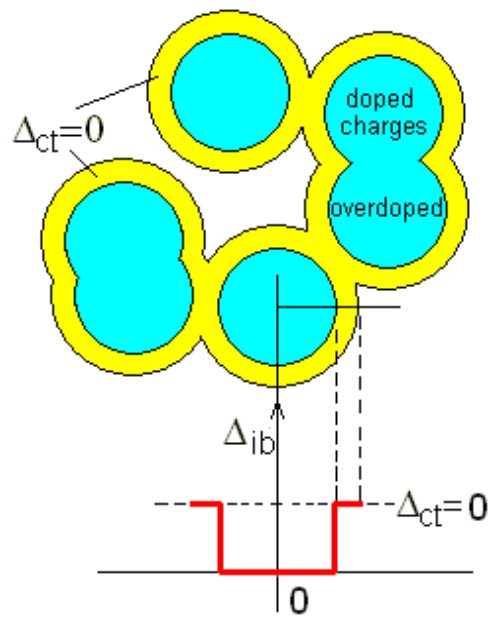

**Figure 5.** The principle of doping in "ignoble" HTSCs. Extra doped charges close the $\Delta_{ib}$ gap in the area where they are localized (green). Under the influence of the extra charges on boundary ions, a layer with $\Delta_{ct}=0$ (yellow) is formed, around the "overdoped" localization area.

Note once again that both in "noble" and "ignoble" HTSCs, the doped charges are localized (except in the ovedoped phase) and their role is reduced only to the formation of areas with $\Delta_{ct}=0$.

### 5. Fermi arcs and the pseudogap in cuprates

It is known that mysterious features in the form of Fermi arcs and the pseudogap are observed on Fermi surfaces of cuprates in some range of temperatures above $T_c$. Here, in the framework of the proposed model, we will show that both Fermi arcs and the pseudogap results from the d-wave symmetry of the order parameter in cuprates and, consequently, they should not be observed in Fe-based HTSCs.

As was mentioned above, pair hybridization on NUC's orbitals with band states takes place in the system under consideration. The width of hybridization $\Gamma$ depends on temperature [3,4] as:

$$\Gamma \approx kT \cdot (V/E_F)^2 \quad (1)$$

(here, $V$ is the one-particle hybridization constant, $E_F$ is the Fermi energy, and $T$ is the temperature).

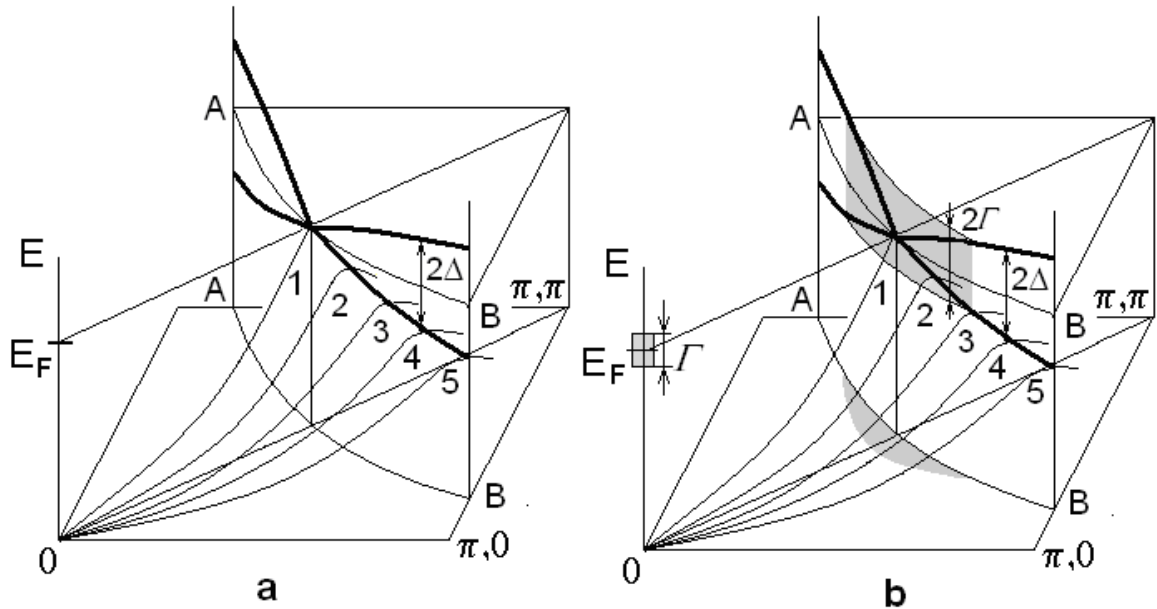

**Figure 6**. Development of Fermi arcs in cuprates: a) $T = 0$, b) $T > 0$. Curves *01–05* are lower branches of the Bogolyubov quasiparticle dispersion curves; the upper branches are not shown. *AB* is the Fermi contour; $\Gamma$ is the width of an NUC pair level. The shaded area around $(\pi/2; \pi/2)$ is the region of momenta of electronic pairs $(k_1, k_2)$ available for transitions to NUCs at a given temperature. The lower (upper) solid curve is the locus of extremum points of lower (upper) Bogolyubov dispersion branches.

The pair hybridization results in transitions of electron pairs ($k_1,k_2$) to NUCs. Each transition is accompanied by the appearance of two quasiparticles -$k_1$, -$k_2$ satisfying the condition $E(\mathbf{k}_1)+E(\mathbf{k}_2) < \Gamma$, where the energies $E(\mathbf{k}_1)$ and $E(\mathbf{k}_2)$ are measured from the Fermi level. At T=0 the NUCs are not occupied (Fig. 6a). As the temperature increases, the region of energies $E$ for which real transitions of electron pairs to NUCs are possible stretches from point ($\pi/2$; $\pi/2$) along the direction of the "crest" of the dispersion, so that a "belt" of height $2\Gamma$, thickness $\Delta k(k)$, and length $L$ along the contour of the FS is formed (Fig. 6b). The arc length $L(T)$ along Fermi contour AB is determined by the condition $\Gamma(T)=\Delta(\mathbf{k})$. The number of such states increases with the temperature as $T^2$ (the shaded area around ($\pi/2$; $\pi/2$) in Fig. 6b).

The NUC occupancy $\eta$ ($0<\eta<2$) is determined by the condition that rates of transitions between the band and the pair-level states in both directions are equal. According to (1), the rate of pair level to the band transitions $\eta\Gamma \propto T\eta$. The rate of the reverse process is determined by the number of band states from which transitions to NUCs is possible and the number of empty NUCs, which means this rate is proportional to $T^2(2-\eta)$. Thus,

$$\eta=2T/(T+T_0) \qquad (2)$$

where constant $T_0$ is independent of the temperature.

The transitions of electron pairs ($k_1,k_2$) to NUCs are accompanied by depairing and result in the formation of Bogolyubov's quasiparticles within a belt of length $L(T)$ and height $2\Gamma(T)$. This conclusion agrees with the results of [6] where Bogolyubov's quasiparticles were observed only around nodes.

Depairing processes should lead to vanishing of the superconducting order parameter around nodes in a arc of length $L(T)$ along the Fermi contour. However, owing to the preservation of coherence in the system, a nonzero order parameter persists on the entire FS excluding the nodes. At the same time, filling of NUCs with real electrons leads to a reduction in the number of NUCs available for virtual transitions of electron pairs. As the temperature increases, the NUC occupancy approaches a critical value $\eta_c$ at which point the superconducting coherence is destroyed and a transition to the normal state takes place. The gap closes along an arc of length L around each nodal direction at the FS due to depairing [7,8]. Meanwhile, along the remaining part of the FS, there still exists a gap (the pseudogap), which corresponds to incoherent pairing [9].

With decreasing doping level there appear Cu ions that do not belong to clusters with $\Delta_{ct}=0$. Such an ion can be thought of as a defect introducing an extra positive potential $\sim\Delta_{ct}$. In the one-dimensional problem, as shown in [10], in the presence of such defect an upper state becomes split off from the band and localized in the vicinity of the defect. In our two-dimensional case, the number of split-off states will depend on the direction of $\mathbf{k}$. As a function of angle, the number of split-off states increases with increasing contribution from Cu orbitals; i.e., this number is the largest for states in the direction of Cu–O bonds. As the number of such defects increases, this leads to the formation of an insulating gap over the FS region from points ($\pm\pi,0; 0,\pm\pi$) towards the nodal directions [11]. The superconducting gap persists only in the FS region adjacent to the nodes, forming islands in the $k$ space [12].

## 6. Fluctuation effects in cuprates (by example YBCO).

According to [13], all Cu ions in $CuO_2$ planes in $YBa_2Cu_3O_7$ belong to NUCs and each $CuO_2$ plane contains a percolation cluster of NUCs. In $YBa_2Cu_3O_{6+\delta}$ with $\delta<0.8$, the percolation cluster of NUCs breaks into finite NUC clusters whose average size decreases with the doping level. In these conditions, the role of the fluctuations in the NUC occupancy increases significantly. According to the suggested model, a transition from the superconducting to the normal state is related to the disappearance of phase coherence taking place as the NUC occupancy approaches the critical value. Thus, whenever a fluctuation causes a decrease in the NUC occupancy, conditions for the restoration of superconducting coherence occur, which can result in "switching-on" of the superconductivity in the temperature range $T^*>T>T_{c\infty}$ (here, $T_{c\infty}$ is the equilibrium value of $T_c$ for an infinite NUC cluster). On the other hand, fluctuation-related increases in the NUC occupancy lead to the disruption of coherence and to "switching-off" of the superconductivity for $T_c<T<T_{c\infty}$. Large fluctuations in the NUC occupancy, corresponding to considerable deviations of $T^*$ and $T_c$ from $T_{c\infty}$, are possible in underdoped samples, where no infinite cluster exists and NUCs are arranged into finite clusters. As the doping level is reduced, the average size of these clusters decreases and relative fluctuations in the NUC occupancy in these clusters grow (i.e., $T^*$ increases and $T_c$ decreases).

In the context of the suggested model, dependences of $T^*$ and $T_c$ on the cluster size can be determined in the following way. We suppose that, for $\delta < \delta_c$, NUCs form finite clusters of some average size $S(\delta)$, and the sample represents a medium, where superconductivity of the entire system appears due to the Josephson coupling between superconducting clusters. We measure the size $S$ of a cluster by the number of Cu sites it contains. Consider a cluster in the $CuO_2$ plane containing a number of NUCs. Then, according to (2), the number of electrons at NUCs in the given cluster at temperature $T$ equals $N=TS/(T+T_0)$. Owing to fluctuations, this number may vary by $\pm\sqrt{N}$. The condition for fluctuating "switch-on" ("switch-off") of superconductivity in the cluster at temperature $T^*$ ($T_c$) can be written out as $N(T)\pm\sqrt{N(T)} = N_c$, where $N_c$ is the number of electrons at NUCs in the cluster for $T=T_{c\infty}$. Thus,

$$TS/(T+T_0)\pm(TS/(T+T_0))^{1/2}=T_{c\infty}S/(T_{c\infty}+T_0). \tag{3}$$

Solving equations (3), we can find the dependences of $T^*$ and $T_c$ on the cluster size $S$ (Fig. 7). Then, relying upon the data on the statistics of finite NUC clusters as a function of the doping level $\delta$ (e.g., in YBCO), we can determine the dependences $T^*(\delta)$ and $T_c(\delta)$ [13], the result being in excellent agreement with the experiment.

Thus, in the region between curves $T_c(\delta)$ and $T^*(\delta)$, clusters fluctuate between the superconducting (coherent) and normal (incoherent) states. The number of NUC clusters being in the superconducting state at a given moment, as well as the lifetime of this state, increase with decreasing temperature. The experimentally measured value of $T_c(\delta)$ has the meaning of a temperature corresponding to the appearance of a percolation cluster of Josephson-coupled superconducting clusters of NUCs. It is evident, however, that, in a certain range of temperatures $T_c(\delta)<T<T^*(\delta)$, sufficiently long-lived and sufficiently large superconducting clusters will be present. In these clusters, the Nernst effect and giant diamagnetism can be observed at $T>T_c(\delta)$ [14,15]. The above discussion suggests that manifestation of these anomalies is not directly caused by the existence of the pseudogap, but rather results from the presence of fluctuating coherent superconducting clusters in the sample.

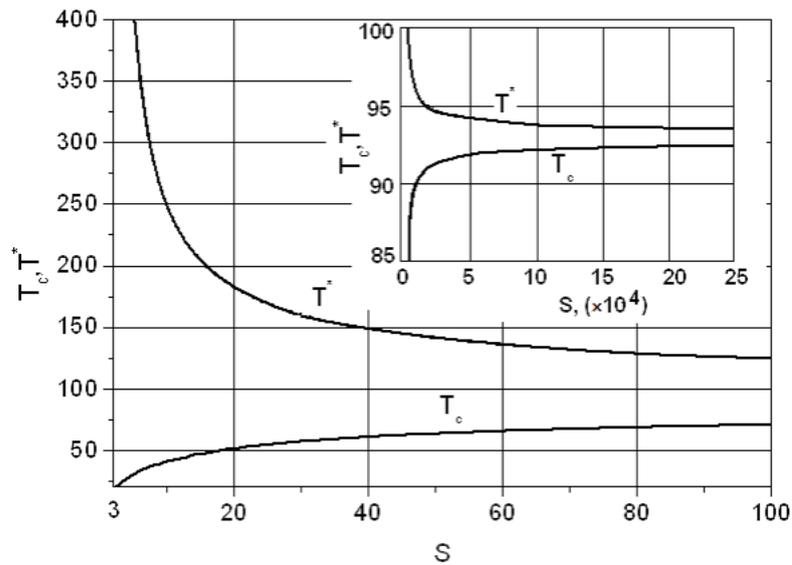

**Figure 7.** Dependences of temperatures $T^*$ and $T_c$ on the cluster size $S$ for $S<100$. Inset: the same dependences for $S<2.5\times10^5$.

It should be noted that, although Fe-based HTSCs lack a pseudogap, similar fluctuation effects caused by switching of small clusters from superconducting to normal state and back can take place in strongly underdoped samples.

**Conclusion**

Thus, we have suggested a qualitative model describing the ground state and the mechanism of superconductive pairing in Cu- and Fe-based HTSCs. In this model, doping by localized charges (as well as physical or chemical pressure) is supposed to be responsible for the closing of the gap between the occupied anionic band and unoccupied states of the cation band and for the formation of the exciton-electronic band of unusual nature. The resulting HTSC ground state is strongly correlated insulator with Fermi surface, where the incoherent electron transport is impossible but coherent superconducting transport is possible because the band is not fully occupied.

It is shown also that such electronic system is inherently predisposed to superconductive pairing because each pair of nearest cations acts as a two-atom negative-U center. The nature of Fermi arcs and mechanism of pseudogap are considered. It is shown that both of these features result from *d*-wave pairing and therefore have to be observed only in cuprates. We believe that the considered ground state is common for various families of HTSCs including cuprates, pnictides, selenides, bismutates and probably some other.